# The evolving of Data Science and the Saudi Arabia case. How much have we changed in 13 years?

Igor Barahona. Department of Information Systems and Operations Management. Business School. King Fahd University Petroleum and Minerals.

### Abstract

A comprehensive examination of data science vocabulary usage over the past 13 years in this work is conducted. The investigation commences with a dataset comprising 16,018 abstracts that feature the term "data science" in either the title, abstract, or keywords. The study involves the identification of documents that introduce novel vocabulary and subsequently explores how this vocabulary has been incorporated into scientific literature. To achieve these objectives, I employ techniques such as Exploratory Data Analysis, Latent Semantic Analysis, Latent Dirichlet Analysis, and N-grams Analysis. A comparison of scientific publications between overall results and those specific to Saudi Arabia is presented. Based on how the vocabulary is utilized, representative articles are identified.

This work evidence a notable absence of data science research related to sustainability and renewable energies. This gap is a significant concern, given the increasing global focus on environmental preservation and green energy sources.

## 1. Introduction

The number of individuals, businesses and organizations that are "always-connected" to the Internet has grown exponentially over the past decade. This phenomenon has generated massive amounts of data and gave birth to the term “big-data”. According to SAS (2022), the term "big-data" is used to denote large volumes of data, both structured and unstructured. Moore's law states that the number of transistors included in an integrated circuit doubles every two years (Moore, 1965), accurately describes the growth in the number of devices connected to the Internet. According to Van der Aalst (2016), digital storage capacity and the number of pixels per unit area have also grown exponentially during the last decade. The foregoing generates a panorama abundant in data which requires innovative methodologies in order to transform such big-data into useful information for decision-making processes. The IDC (2022) states that by the year 2023, around 75% of G2000 companies will pledge to ensure technical equality for a workforce intentionally designed to function both remotely and in-person, facilitating seamless collaboration in real-time.

Given this context, a new discipline called "Data Science" emerged. Contemporary digital societies face new challenges, which cannot be solved through traditional scientific methods. For instance, it is not possible to extract useful information from large amounts by applying the well-known traditional mathematical methods, which were suitable three decades ago. In order to successfully tackle with big-data, the combination of mathematical methods and computational sciences around a specific application domain is required. In this way, data science emerges as a fusion between computational science, mathematics and statistics around a specific domain. Skiena (2017) states that data science is at the intersection of computational sciences, statistics and mathematics, which establish synergies to investigate a particular phenomenon.

On the other hand, Saudi Arabia is the biggest country in Middle East. With a Gross Domestic Product (GDP) around 8% for 2022, Saudi Arabia is one of the economies with faster growing after COVID crisis and one of the most digitalized economies in Middle East (SASt-a, 2023). According to the Saudi Authority of Statistics (SASt-b, 2023) the population was 34,110,821 on year 2021. In 2019 the 24% of adults in Saudi Arabia (25-64 years old) had attained a tertiary qualification (OCDE, 2022). The tertiary Saudi system is mainly composed by 80 universities which are distributed across all regions of the country (MOE, 2023). A search on SCOPUS yields to 373,853 papers which are authored by researchers affiliated to Saudi Universities. Among the main research disciplines conducted in Saudi Arabia are: medicine, engineering and computer science. Around the 50% of these papers were published during the last five years (SCOPUS,2023). Considering the mentioned trends, it results pertinent to dig deeper on how these tendencies are related with the data science evolving. Under this framework, the general objective for this work consists of investigating how the data science is evolving during the last 13 years, additionally three specific objectives stated as it shown below.

- Given a collection of 16,018 scientific abstracts that contain the words “data science” either on the tittle, abstract or keywords, identify representative articles for the data science field.
- Propose main topics around data science for the investigated period.
- Conduct a comparison of scientific production between the overall results and Saudi Arabia.

To achieve these objectives, the manuscript comprises five sections. A literature review that includes formal definitions for data science, how this discipline emerged and the most remarkable authors, on next section is provided. The characteristics of the investigated dataset and the natural language processing techniques that are applied on our analysis, on the third section are explained. The obtained results are reserved for the fourth section. The conclusions are provided on the last section.

## 2. Literature Review

Donoho, (2017) and Cao, (2017) agreed that Tukey (1962) was one of the first of shaping a construct for data science in the following way: *statistics must be complemented, among other things, with procedures for analyzing data, techniques for interpreting the results of such procedures, with ways of planning the collection of data in order to make their analysis easier, more precise and more exact, in addition to applying all the statistical (mathematical) machinery to data analysis*. Peter Naur, in his book entitled "Concise Survey of Computer Methods", uses the term data science as equivalent in meaning to computational science, mainly focusing on illustrating the relationship between the natural sciences and the use of data (Belzer, 1976). In 1998, during the annual meeting of the International Association of Classification Societies, Hayashi (1998) introduces the concept of "*Dēta no bunseki-teki chōsa*", which can be translated from Japanese to English as the task of "performing an exploration data analytics". Hayashi defines data science as a discipline that clarifies the understanding of a phenomenon through data analysis and carefully designed experiments.

Jeff Wu, a professor in Georgia Tech's School of Industrial Engineering and Systems Engineering, suggested that statistics could be renamed data science, this latter emphasizing data collection, mathematical modelling, and computer systems (Wu, 1997). One of the first authors that coined the concept "data science" in the form that it is understood nowadays was Willian Cleveland. In the article entitled "Data Science: An Action Plan to Expand the Technical Areas of the Statistics Field" he suggests that data science is a domain mixed between mathematics, statistics and computer science Cleveland (2007). Additionally, the author proposes a curricular plan for the development of data science, which is suitable to be applied in governments, laboratories, organizations and universities. In 2002 the "Journal of Data Science" showed up, as one of the first scientific journals specialized in this area. It was followed by others such as "International Journal of Data Science and Analytics", "Data Science and Management" and "Data Science and Engineering", among others.

In 2012, data science was recognized as one of the fastest growing and expanding job areas in the world. The "data scientist" job was highlighted as one of the best paid in the United States (Davenport & Patil, 2012). These authors highlight the qualities and responsibilities that professionals dedicated to data science must possess, as well as their importance in organizations and government entities. In a survey distributed among 11,514 members of the American Statistical Society in 2020, 82% of respondents said they were working in areas such as applied statistics and data science. 92% said that if a student or young person asked them for advice on how to select a university or career, they would encourage them to study statistics or data science. On the contrary, only 7% mentioned that they would discourage students or young people from studying a career related to these areas (ASA, 2020).

Data Science is an exponential growing field in Saudi Arabia. By collecting and analyzing data, government, private enterprises and nonprofitable organizations at Saudi Arabia are willing to improve services for citizens, obtain better understanding of consumer behavior's, leading to more efficient operations and better decision-making. According to the "Saudi Arabia Big Data and Artificial Intelligence Market - Growth, Trends, COVID-19 Impact, and Forecasts" report, Saudi Data Science market was valued around USD 165 million on 2020, and it's expected to reach USD 892 million by 2026 and consequently become the biggest and most valuable data science market in Middle East (R&M, 2023). According to Sherif (2022) there are 48 official programs specialized on Data Science available in Middle East of which 16% are offered by Saudi Universities. The above posits this country on top one related with the data science programs offered in the region. Princess Nourah bint Abdulrahman University, which is the largest female university in the world, has one of the oldest degrees in the field when the master's in computing and data science was launched on 2018 (PNU, 2023).

Along the COVID pandemic, lockdowns and social distance regulations were implemented to reduce the spread of the virus and avoid the medical system collapse. All events considered as nonessential were carryout digitally. This led to unseeing growth on activities as online banking and shopping, streaming services, remote working, and education. Governments collected digital data for monitoring the pandemic evolving and make decisions based on scientific evidence. Among the most remarkable initiatives for collecting COVID data are Covid-19 Monitoring Dashboard by Johns Hopkins University (JHU, 2023), World Health Organization Covid- 19 Dashboard (WHO, 2023) and Covid-19 Global Tracker by Our World in Data (OWD, 2023). Saudi Arabia government created an innovative app for collecting COVID data and managing curfews. Through Tawakkalna app, the government granted electronically moving authorizations to individuals by keeping records which were useful for better decision making (Tawakkalna, 2023).

Based on the mentioned before, data science perspective for post pandemic age is promising. According to the IDC (2022) digital economies will keep growing and accumulating more data. Governments and companies will keep investing on high scope data science projects as ChatGPT, (OpenAI, 2023) and DeepMind AI (Google, 2023). The US Labour of Statistics states that the demand for data scientist will increase by 15% by this decade, which is above the average for other economic activities in the country (USL, 2023). As the need for data science continues to grow, individuals able to demonstrate skills for analysing data will find promising opportunities in the labour market.

# 3. Methodology

As it is illustrated on Figure 1, the proposed methodology comprises six steps. Starting with an explanation of the procedures related to data collection, going through the introduction of the computational tools and closing with the statistical methods applied along the project. In further lines, a detailed explanation is each step is provided.

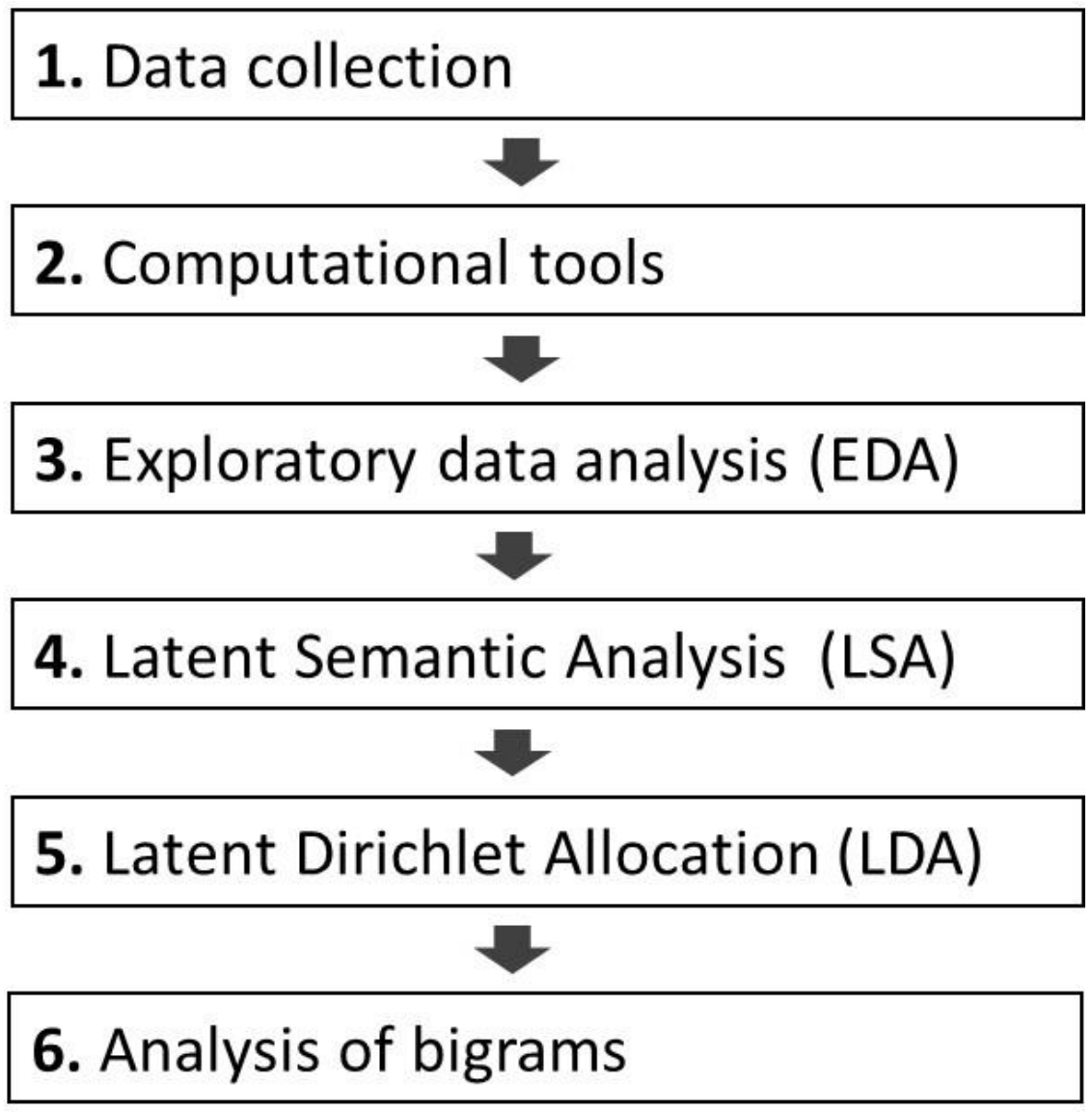

**Figure 1**. Flowchart with the proposed methodology

## 3.1 Data collection

A search was conducted in the SCOPUS (2023) digital library by retaining all the documents that fulfill the following search criteria: all documents must include the words "*data science*", either in the title, abstract or keywords. On Barahona., et. al (2018) it was conducted related research, on which a total of 1,169 were investigated. This work was taken as reference and a total of 16,018 documents were collected, which cover a period between 2007 and 2022. As far as I know, there is no existing research in the literature that investigates the vocabulary of data science from a text mining perspective while spanning 13 years of scientific research.

## 3.2 Computational tools

The analysis was carried out under the R programming environment, version 4.0.0 (Already Tomorrow) released for download in April 2023. The following specialized text mining libraries “tidytext”, “factoextra”, “topicmodels” were used (CRAN, 2022). A complementary analysis was conducted on Python 3.0, by using the following libraries: Pandas, NLKT, Scikit-Learn, Matplotlib.

### 3.3 Exploratory Data Analysis

In this step, a preliminary examination of the dataset was conducted to identify key characteristics, select the appropriate statistical methods for the analysis, and generate hypotheses regarding the possible relationship between the variables being investigated. This is a crucial step since it allowed us to retrieve the first insights for the corpus and then further prepare the strategy for a deeper analysis. Also, in this stage the "stopwords" were removed, which mainly consist of articles, pronouns and prepositions, among others. The algorithm proposed by Schofield, Magnusson & Mimno (2017) was adapted for these purposes.

### 3.4 Latent Semantic Analysis (LSA)

The Correspondence Analysis (CA) is a well-known statistical method for dimensionality reduction proposed by Jean-Paul Benzécri in the 1960s, which has been widely used to investigate datasets with great complexity or size. The starting point for carrying out an CA is a Document-Term-Matrix (denoted as ***X***), with dimensions ***n*** x ***p***, where ***n*** represents the number of observations (rows) and ***p*** represents the variables (columns). In Figure 2, a representation of the matrix ***X.***

$$
\begin{array}{c|cccc|c}
 & \mathrm{B}_1 & \mathrm{B}_2 & \cdots & \mathrm{B}_J & \\
\hline
\mathrm{A}_1 & f_{11} & f_{12} & \cdots & f_{1J} & f_{1\cdot} \\
\mathrm{A}_2 & f_{21} & f_{22} & \cdots & f_{2J} & f_{2\cdot} \\
\vdots & \vdots & \vdots & \ddots & \vdots & \vdots \\
\mathrm{A}_I & f_{I1} & f_{I2} & \cdots & f_{IJ} & f_{I\cdot} \\
\hline
 & f_{\cdot 1} & f_{\cdot 2} & \cdots & f_{\cdot J} & n
\end{array}
$$

**Figure 1.** Structure for a Document Term Matrix (DTM)

For the Word Document Matrix (MTM) $f_i = \sum_j f_{ij}$ represent the marginal frequency of $\mathrm{A}_I$. On the other hand, $f_j = \sum_i f_{ij}$ represents the marginal frequency for $\mathrm{B}_J$. Note that each category $\mathrm{A}_1, \dots, \mathrm{A}_I$, corresponds to the values contained on the vector $\boldsymbol{a} = (a_1, \dots, a_I)$. Each element of $\boldsymbol{a}$ is given by the absolute frequency at the respective position of the DTM. Similarly, for the columns $\mathrm{B}_1, \dots, \mathrm{B}_J$, that correspond to values of the vector $\boldsymbol{b} = (b_1, \dots, b_J)$.

Let ***U*** and ***V*** be composite variables, which are defined as $\boldsymbol{U} = \boldsymbol{X}a$ and $\boldsymbol{V} = \boldsymbol{Y}b$, respectively. Then, the correspondence analysis consists of finding the linear combination for the vectors $\boldsymbol{a}$ and ***b***, which maximize the correlation between the composite variables ***U*** and ***V***. Then, the Singular Value Decomposition (SVD) for the matrix ***X*** is obtained as it is shown in formulation (1).

$$\mathbf{D}_a^{-\frac{1}{2}}(\mathbf{X} - ab')\mathbf{D}_b^{-\frac{1}{2}} = U\mathbf{D}_\lambda V' \qquad (1)$$

In the context of text mining and natural language processing, this methodology makes it possible to quantify relationships between words and documents, based on their semantic connotation. The foregoing through the comparison of the *row-categories* on the one hand, and the *column-categories* on the other, by taking eigenvalues obtained with formulation (1) as reference. Two or more words with close eigenvalues will have similar semantic meaning.

The above applies for documents. From two or more documents with similar eigenvalues, it is inferred that they are using similar vocabulary, or embracing on similar topics. By introducing the year of publication variable as supplementary to the analysis, it is possible to identify the evolution of the vocabulary over time. The year of publication will be characterized by the words or documents that have close eigenvalues. Subsequently, through the matrices obtained with the SVD, it is possible to generate visualizations colloquially called "*clouds of words*" or "*clouds of documents*".

### 3.5 Latent Dirichlet Allocation (LSA)

This is an unsupervised model proposed by Blei., et. al. (2003) that finds wide applications in natural language processing and text analysis. Its primary purpose is to unveil latent topics within textual datasets by leveraging the Dirichlet distribution, which is a generalization of the Gamma distribution. LDA is classified as a generative model because it seeks to characterize the underlying patterns within a corpus. More specifically, given parameters $\boldsymbol{k}$ and $\boldsymbol{\alpha}$, the Dirichlet distribution is employed to infer how topics are generated from a collection of documents, as illustrated bellow.

$$p(\theta|\alpha) = \frac{\Gamma(\sum_{i=1}^{k} \alpha_i)}{\prod_{i=1}^{k} \Gamma(\alpha_i)} \; \theta_1^{\alpha_1 - 1}, \ldots, \theta_k^{\alpha_k - 1} \qquad (2)$$

According to formulation (2), $\boldsymbol{k}$ represents the number of latent topics on the corpus and $\boldsymbol{\alpha}$ is the Dirichlet prior referring to the topic mixture for each document included in the dataset. If the parameter $\boldsymbol{\alpha}$ is interpreted as the proportion of topics observed in each document, then $p(\theta|\alpha)$ represents the probability that the document $\boldsymbol{\theta}$ contains the topic $\boldsymbol{\alpha}$.

### 3.6 Analysis of bigrams

Given a corpus, this step involves examining pairs of consecutive words. A bigram is defined as a sequence of two adjacent words. In this work, bigrams were calculated to capture the context of words around the data science concept. A collection of 16,018 of documents was taken as input to obtain the frequency table for bigrams. Subsequently, those bigrams with observed frequency equal of higher than 150 were retained for generating visualizations that allowed us to have a better understanding of the data science vocabulary.

## 4. Results

### 4.1 Exploratory Data Analysis

The number of publications per year for the period from 2009 to 2022 is presented in blue color in Figure 2. To estimate the number of publications related to data science for years 2023 and 2024, the following second order model was adjusted: $y(x) = 3.13x^2 - 1.25^5x + 1.27^8$ through the method of least squares. With an accuracy of $R^2$=0.91 and p-value=$1.7^{-7}$, it is forecasted that the number of will be 4,726 and 5,540 for years 2023 and 2024 respectively.

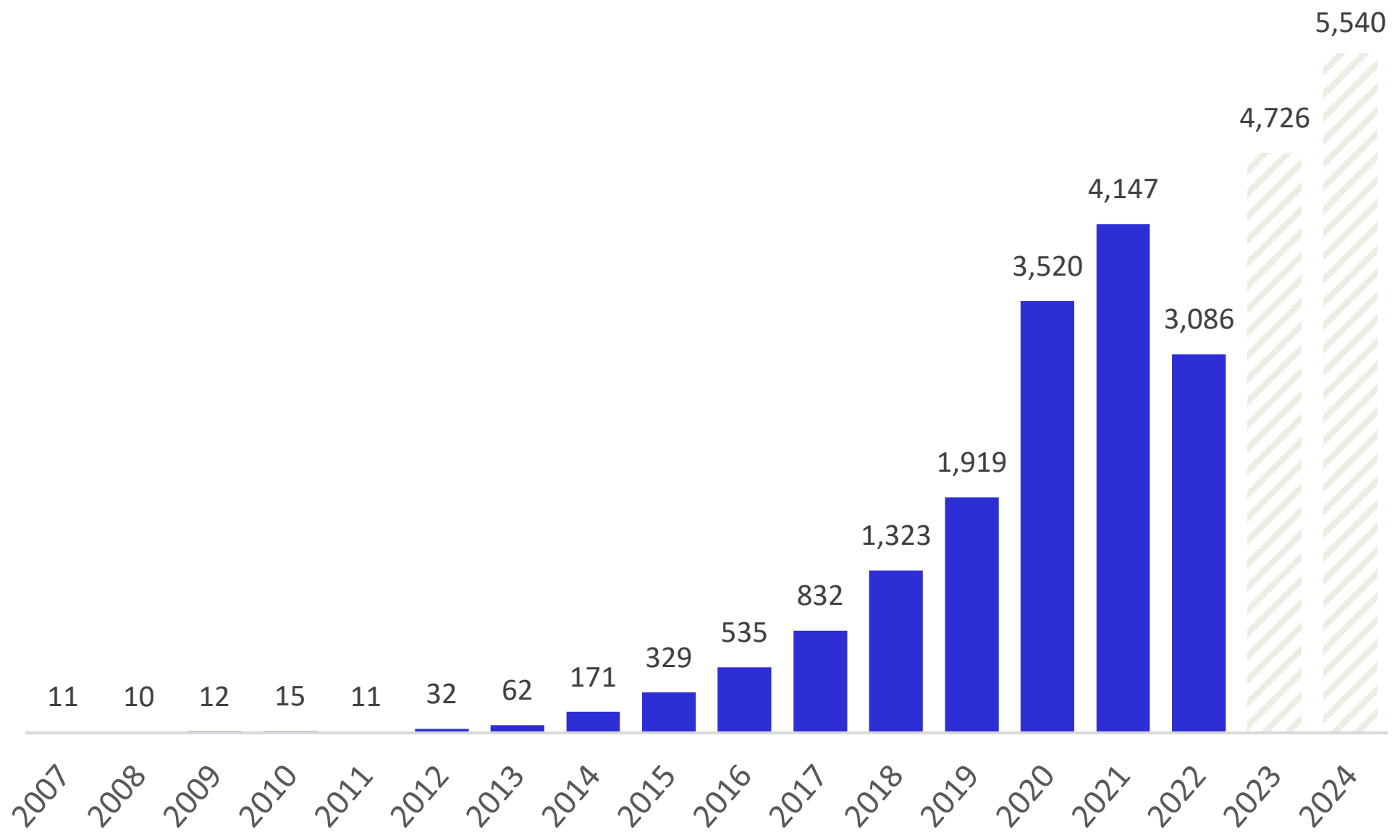

**Figure 2**. Number of publications per year

Words such as “learning” (8,996), “analysis” (6,980), “model” (6,818), “paper” (6,591), “methods” (5,186) and “machine” (5,087) appear as the most frequent in the database under study. Through this first approximation, we can get a general idea of some of the predominant topics for data science during the last decade, such as “machine learning” or “machine learning methods”. The thirty words that appear in Figure 3 and their respective frequencies represent 10.1% of the total volume of the vocabulary.

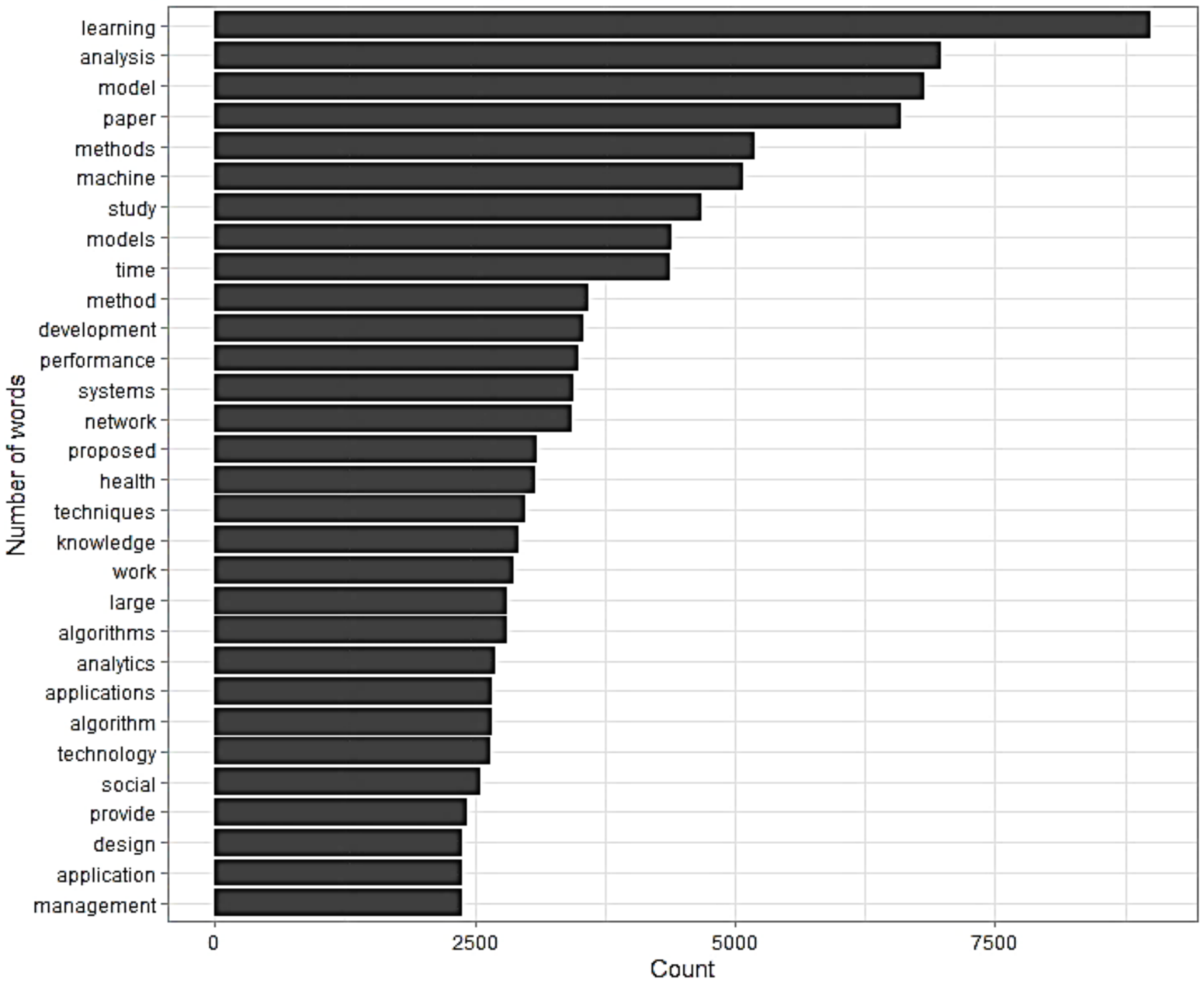

**Figure 3**. Most frequent words in the topic of data science. Period from 2007 to 2022

Regarding the type of publication, 51% of the total publications on this topic correspond to "Conference Proceedings". The rest is divided between "Research Articles" (37%), "Book Chapters" (7%), "Conference Reviews" (3%), "Books" (2%) and "Editorials" (1%). In this way, conference proceedings are the preferred type of publication for academics and scientists who carry out research on topics related to data science.

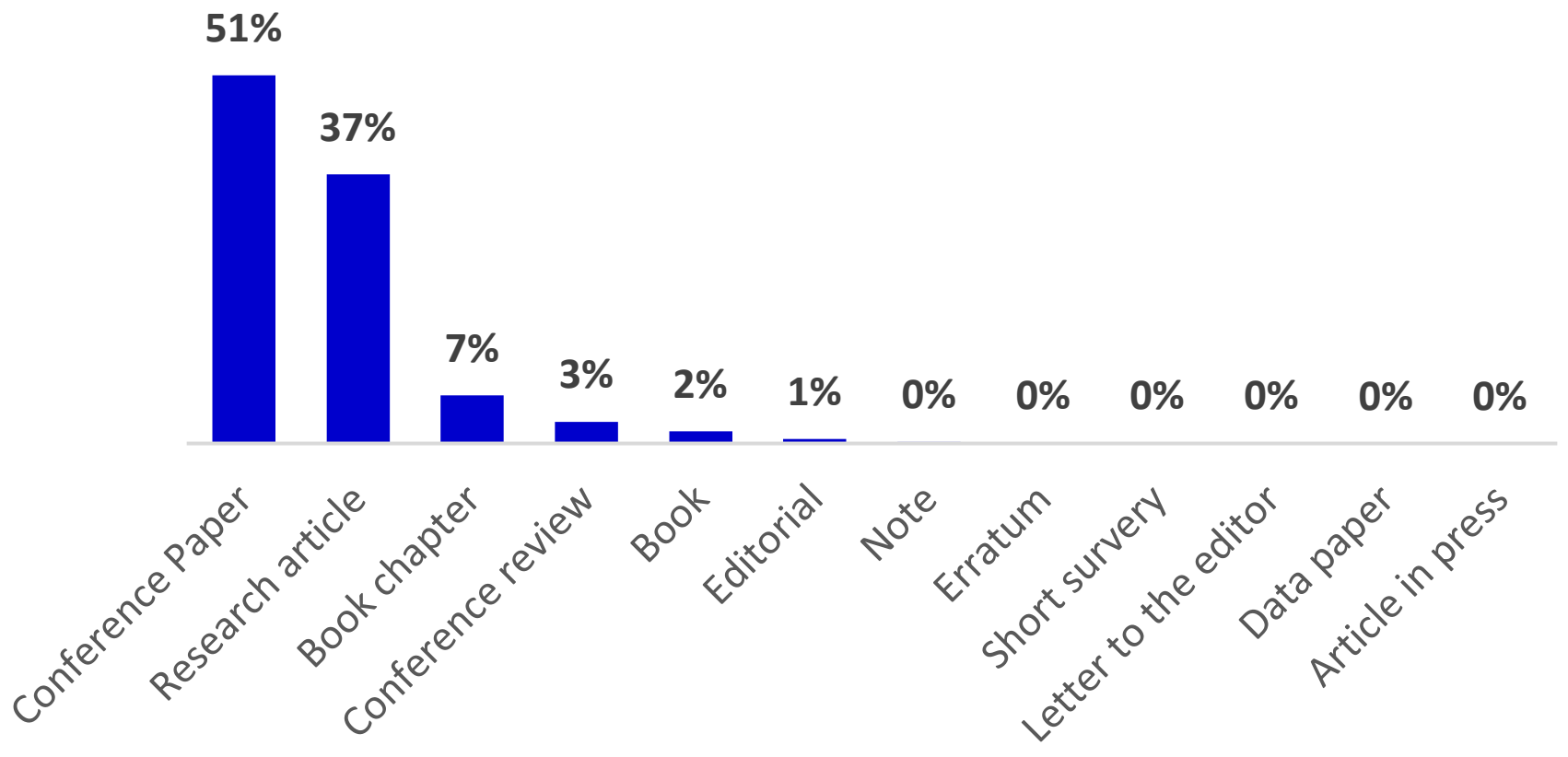

**Figure 4**. Type of publication for "Data Science". Period from 2009 to 2022

Articles published in conferences are generally the result of participation in an annual meeting of a union or association. Usually, these documents are short, precise and with a reduced number of pages. Subsequently, a selection of these documents is published, which is called "conference proceedings". On the other hand, the articles are generally published in specialized journals on a particular topic. Unlike conference proceedings, journal articles can be published without the requirement of making an oral presentation of the work. In general, articles published in recognized journals can have a greater impact than conference proceedings. However, the above may be different depending on each research area.

## 4.2 Latent Semantic Analysis

As previously explained on the methodology, a Document-Term Matrix (DTM) of order with dimensions ***n***=16,081 and ***p***=1,000 was generated.  While refers ***n*** to the number of documents included in the study, ***p*** is an arbitrary parameter that denotes the number of words retained on the analysis. Later, formulation (1) was applied to factorize the DTM into three elements. At first, ***U*** comprises the eigenvalues for the ***n*** documents. Two dimensions were preserved, therefore ***a***=2. Finally, ***V*** refers to the eigenvalues for the words. According to Lebart,. et. al. (1997), two or more elements on matrix ***U*** with similar eigenvalues will be related semantically. Moreover, those articles with the highest eigenvalues can be considered as representative, since they comprise most of the semantic terms included on the corpus. In this way, representative articles summarize the corpus characteristic concepts.

On Figure 5 a visual representation of the Latent Semantic Analysis is provided. Vertical and horizontal axles refer to the first and second components with ***a***=2. Each dot represents a document included in the analysis. For sake of simplicity only the top-five representative articles were labelled. The articles titled "A single program multiple data algorithm for feature selection" (**A8626**) discusses how a Single Program Multiple Data (SPMD) approach is implemented on Minimum Redundancy Maximum Relevance (mRMR) algorithm (Chanduka., et al. 2020). The article titled "Can Statewide Emergency Department, Hospital Discharge, and Violent Death Reporting System Data Be Used to Monitor Burden of Firearm-Related Injury and Death in Rhode Island? (**A4798**) illustrates how data science is applied in Medicine (Jiang., et. al. 2019). The document authored by Lakshmi., et. al. (2022) and titled "An Optimal Deep Learning for Cooperative Intelligent Transportation System" presents and application for data science for transport optimization (**A14959**). In the article titled "Understanding Performance Concerns in the API Documentation of Data Science Libraries", Tao et., al. (2020) address to data science practitioners by discussing the main challenges on developing data science applications (**A7222**). The book chapter titled "Conceptual design principles for data-driven clinical decision support systems (CDSS): Developing useful and relevant CDSS" is considered the most representative among the investigated corpus (**A15093**). In this document, Zikos (2023) discusses design principles for data-driven clinical decision support systems.

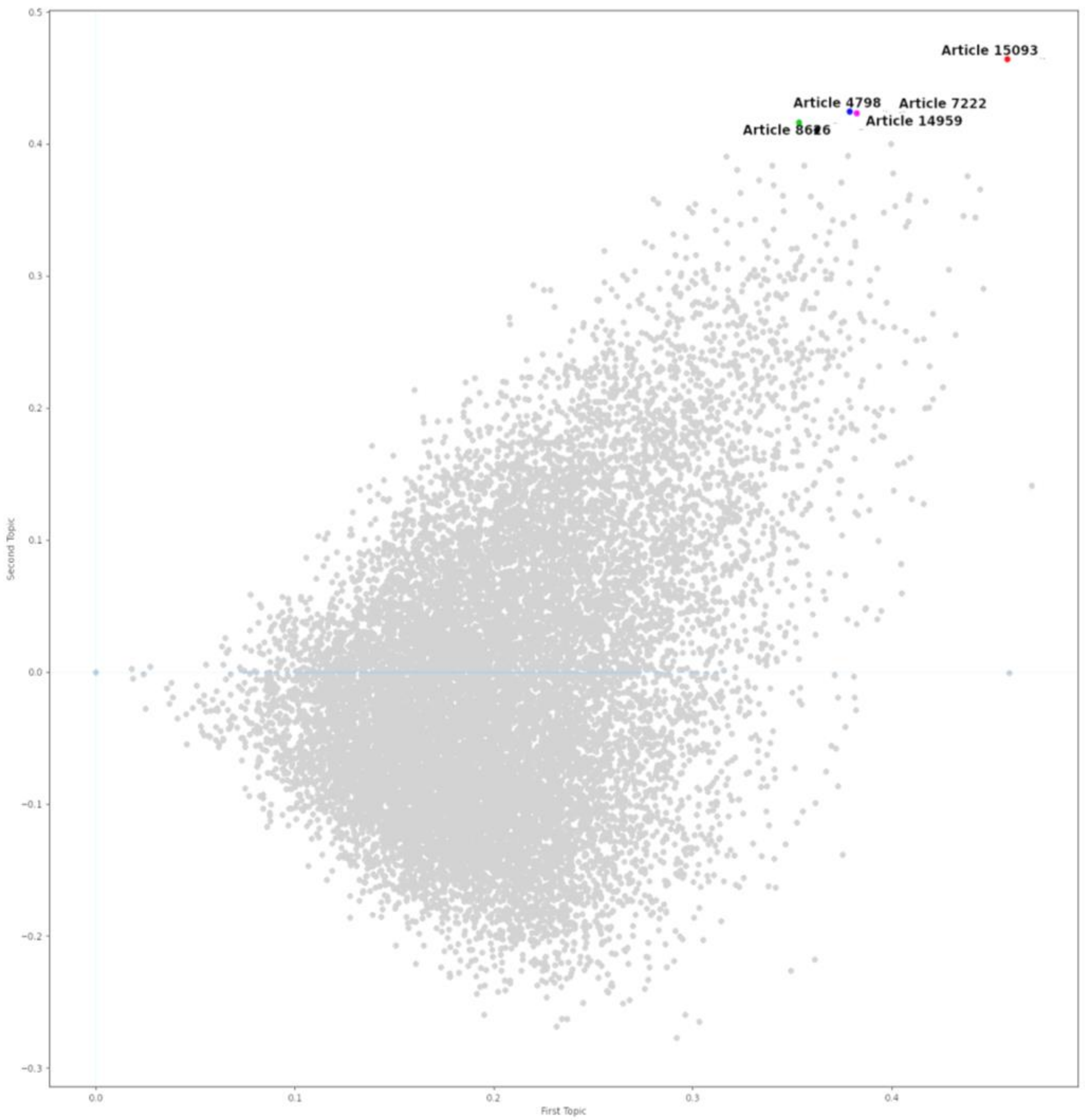

**Figure 5**. Most influential articles for data science filed in the period 2009-2022

## 4.3 Latent Dirichlet Allocation

In this subsection, results for identifying the most relevant topics underlying Data Science literature are provided. LDA assumes that documents are mixtures of topics, similarly, they are also given by a mixture of words. First, each document is allocated to a given topic, then those words contained within the document are identified based on the previously defined distribution of topics. For this work, it was arbitrarily defined the number of topics equal to six, as it is illustrated on Figure 6.

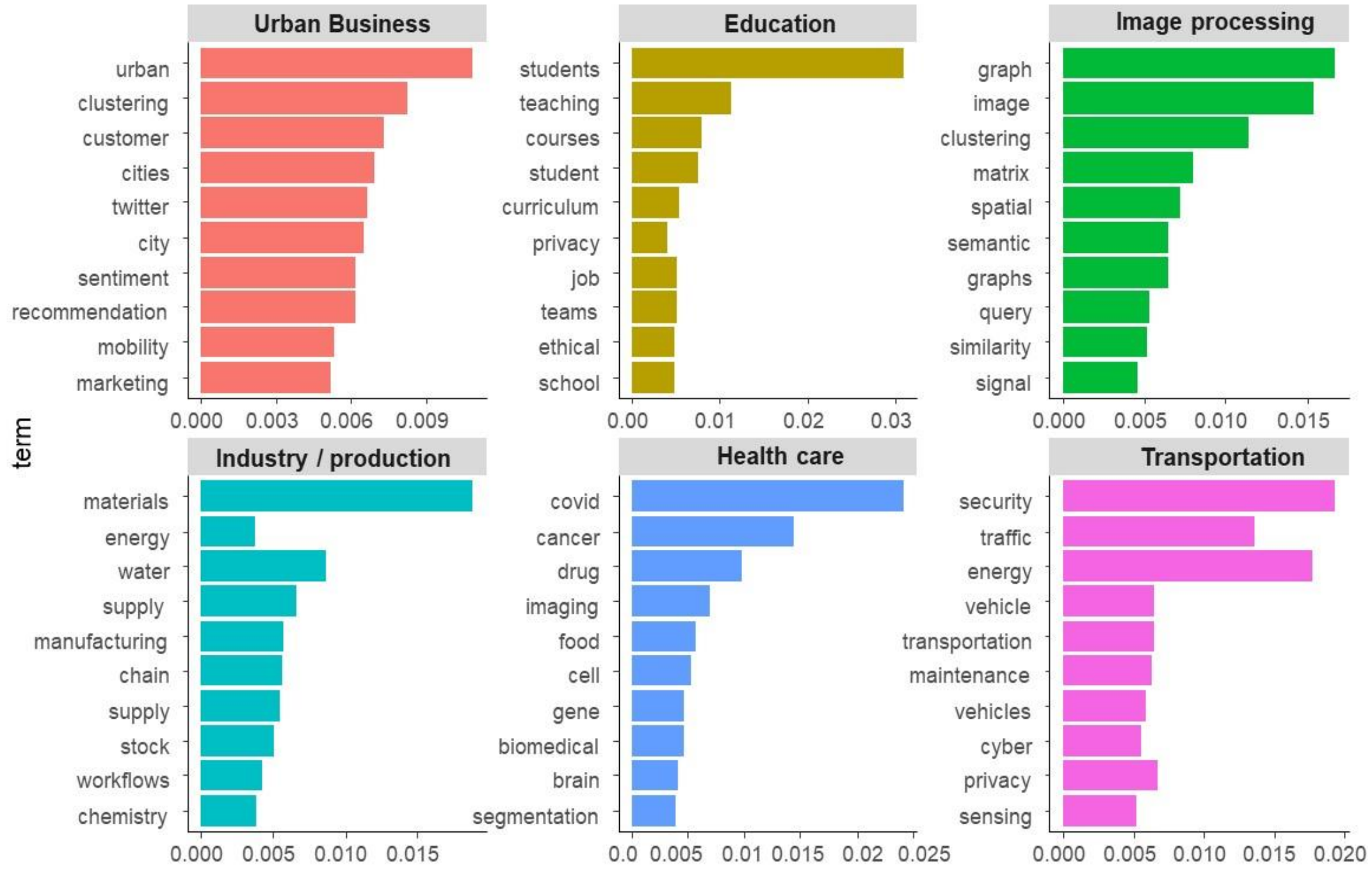

**Figure 6**. Most relevant topics for Data Science during the period 2009-2022

The first topic "Urban Business" comprises the words "urban", "cities", "sentiment" and "marketing". Secondly, "Education" includes the words "students", "teaching", "courses", and "school". From the above it is inferred that an important part of the investigated documents related to data science, take education as domain of application. "Image processing" is an important branch of data science. This topic contains wors as "graph", "image", "clustering", "matrix" and "special". The fourth topic, "Industry / production" includes words as "materials", "energy", "supply", "manufacturing" and "workflows". In fifth place, "health care" comprises words as "covid", "cancer", "drug", "cell" and "biomedical". The above yields supportive evidence in favor of the importance of data science for health care and health advancements. The last identified topic is "transportation", which is characterized by words as "security", "traffic", "energy", "vehicle" and "transportation.

## 4.4 Bigrams Analysis

As it was previously mentioned, bigrams allow us to retain the most frequent terms, and therefore disclose patterns on the bases of the vocabulary is being used. On Figure 7 a visual representation of the bigram analysis is provided. Some of the most frequent bigrams are "machine learning", "artificial intelligence", "deep learning", "decision making", "neural network", "learning algorithm", and "large scale".

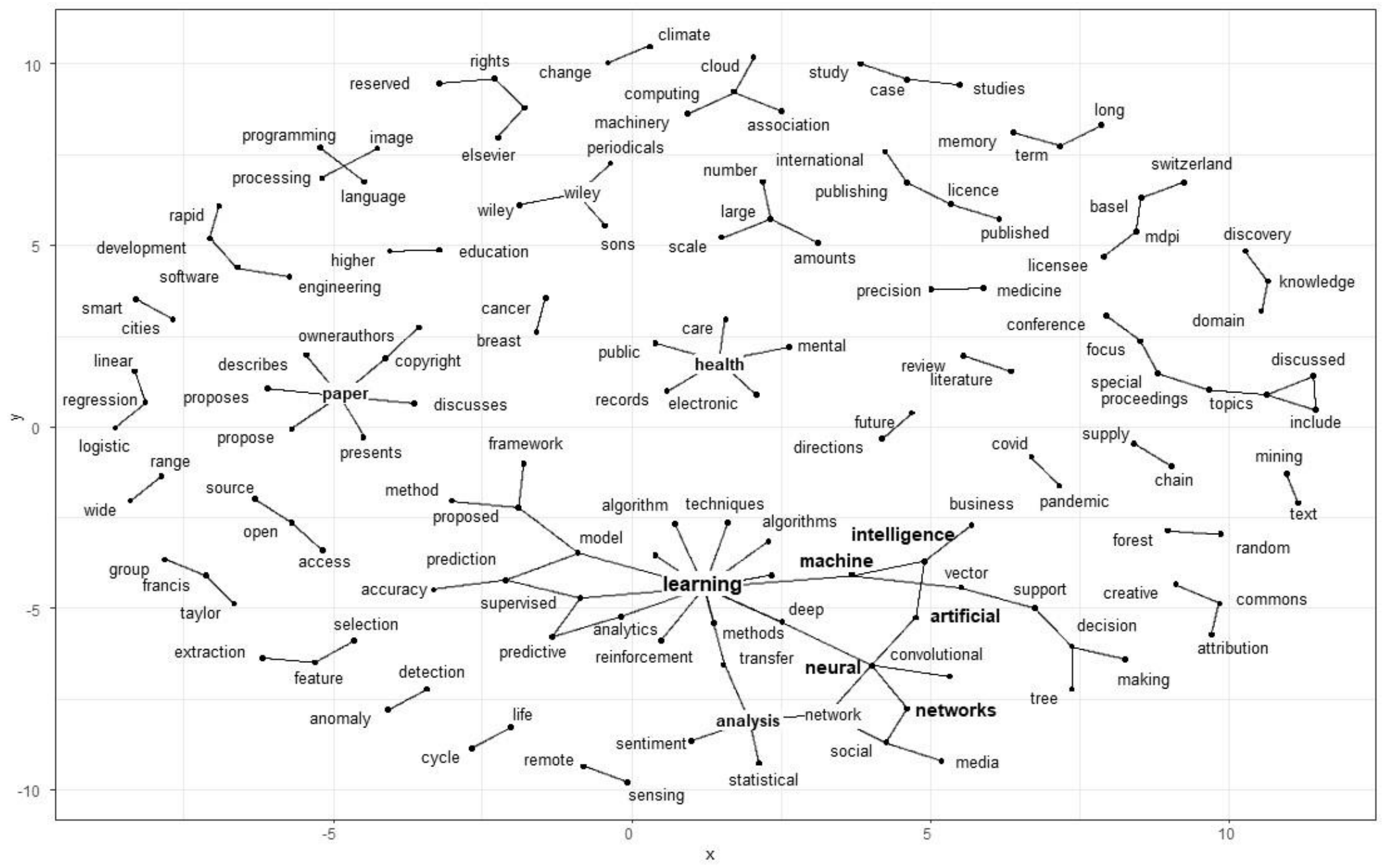

**Figure 7**. Most frequent bigrams for Data Science during the period 2009-2022

Mentioned bigrams are the key concepts on which data science is founded. Machine learning, artificial intelligence, and deep learning are indeed the core of data science. The bigram "large scale" emphasizes the importance of scalability while dealing with big data, which common element of data science. The ultimate purpose for data science consists of improving decision-making processes, as is evidenced in this analysis. Practitioners can use these bigrams as reference while conducting scientific literature searches.

## 4.7 The case of Saudi Arabia

For the last subsection, articles authored by Saudi institutions were filtered. This query reached a total of 233 documents, which correspond to the 1.5% of the investigated corpus. The above contrast with countries as United States that comprise 4,973 documents, representing the 31% or China with 1,941 documents corresponding to 12.1%. A comparison of principal characteristics between Saudi dataset and whole corpus, on Table 1 is presented.

**Table 1.** Comparative statistics for Saudi Arabia

| Descriptive statistics | Saudi Arabia | Corpus |
|---|---|---|
| Number of documents | 233 | 16,018 |
| Number of words | 48,159 | 3,023,613 |
| Average words per document | 204 | 188 |
| Largest document (word count) | 414 | 1,283 |
| Shortest document (word count) | 153 | 139 |

According to the Saudi Arabia vision 2023, the kingdom is focused on strengthening the research in areas as data science and artificial intelligence. However, a comparison with the overall and other nations reveals that the output of scientific research originating in Saudi Arabia and related with data science is significantly lower. To foster greater scientific research output, Saudi Arabia is actively investing in education and research, aiming to cultivate a culture of innovation and knowledge creation, which will likely lead to an increase in scientific publications in the years to come.

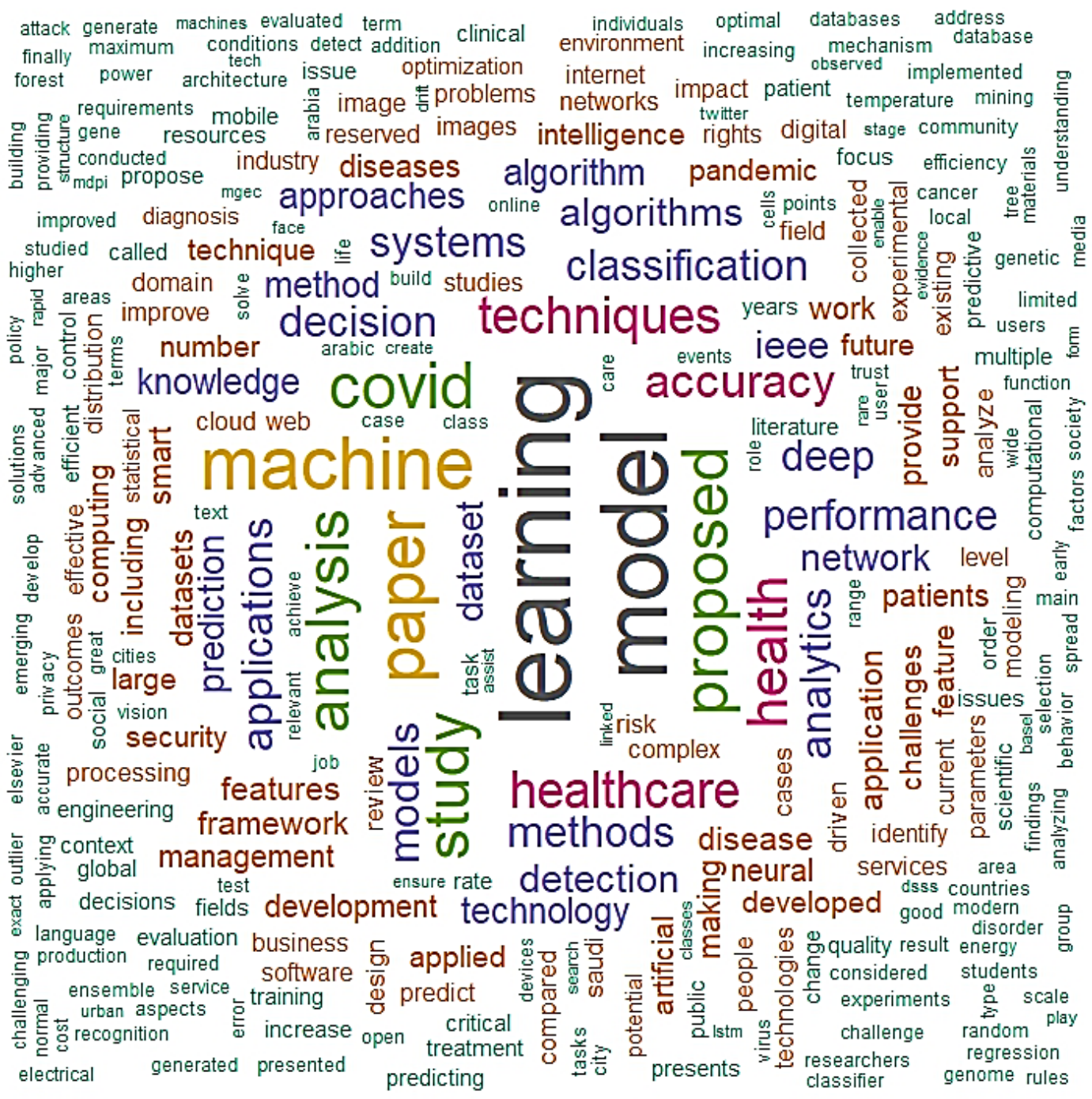

**Figure 8**. Word cloud for articles authored by Saudi institutions. A total of 233 documents.

On Figure 8 a word cloud was prepared with the documents authored by Saudi institutions. A total of 1,000 terms were retained with a frequency equal or higher than 10. Among the most frequent terms are “learning”, “model”, “machine”, implying the Saudi research is focused on machine learning and modelling. "Techniques," "accuracy," "methods," and "models" underscore the emphasis on refining analytical and modeling techniques, particularly in the context of classification and decision-making. "Deep," "network," and "algorithm" evidence that Saudi institutions are applying deep leaning and neural networks. "Disease," "pandemic," and "diagnosis" indicate the dedication of Saudi institutions for addressing public health challenges. “Business," "industry," and "Internet" highlight the intersection of science and the business world. This word cloud indicates that Saudi research in the field of data science is aligned with the worldwide trends in this area.

## 5. Conclusions

The transition from analogical to digital societies has generated massive amounts of data. This tendency has never seen before in the history of the humankind, and it is expected to continue in the coming years. Derived from this, contemporary organizations face new challenges as processing and analyzing big data in order to improve decision making. Data science emerged was a discipline focused on transforming raw data into useful information for improving decision making processes. Under this perspective, this work analyses 16,018 documents which contain the term “data science” on either the title, abstract or key words. The above derived on a corpus that contains 3,023,613 words. To get insights from this corpus, this works applies exploratory data analysis was conducted, followed a Latent Semantic Analysis, Latent Dirichlet Allocation and analysis of bigrams.

Based on the model presented in subsection 4.1, an upward trend in the number of publications related to data science in the coming years is anticipated. It's worth noting that 51% of the documents pertain to conference proceedings, which contrasts with other disciplines like business management or operations research, where journal articles predominate. As per the Pareto Principle, roughly 80% of the effects stem from 20% of the causes, and in this case, the top six most frequent words, namely "learning," "analysis," "model," "paper," "methods," and "machine," collectively account for 10% of the corpus. Utilizing Latent Semantic Analysis, I've identified the most representative articles, and through Latent Dirichlet Analysis, I've proposed six primary research areas: "urban business," "education," "image processing," "industry/production," "health care," and "transportation," as the main topics within data science. Furthermore, a bigrams analysis has revealed that the principal research themes in data science over the last decade revolve around "machine learning," "artificial intelligence," "neural networks," and "public health." Lastly, a comparison of key characteristics between the Saudi dataset and the entire corpus has unveiled two key findings. While the research areas pursued by Saudi institutions align with the overall results, the fraction of Saudi documents significantly lags behind countries like the United States and China. In response, Saudi Arabia is actively implementing large-scale programs to foster a culture of innovation and knowledge creation.

One notable gap exposed by this study is the absence of data science research related to sustainability and renewable energies. This gap is a significant concern, given the increasing global focus on environmental preservation and green energy sources. Within this context, practitioners, researchers, and decision-makers in the field of data science possess a remarkable potential for generating innovation and research aligned with the pressing global requirements. The adoption of sustainable practices and the transition to renewable energies are central issues for contemporary societies. Consequently, the integration of data science into these domains is a strategic imperative for achieving high-impact results.